\documentclass[aps,prl,reprint,amsmath,amssymb,superscriptaddress,longbibliography]{revtex4-2}

\usepackage[bookmarks=false,linkcolor=NavyBlue,urlcolor=NavyBlue,colorlinks,citecolor=NavyBlue]{hyperref}
\usepackage{graphicx}
\usepackage{amsmath}
\usepackage[svgnames]{xcolor}
\usepackage{enumerate}
\usepackage[normalem]{ulem}
\usepackage{orcidlink}
\usepackage{comment}

\graphicspath{{./Figures/}}

\renewcommand{\vec}[1]{\mathbf{#1}}

\newcommand{\figref}[2]{\hyperref[#1]{\ref{#1}(#2)}}
\newcommand{\figsref}[2]{\hyperref[#1]{\ref{#1}#2}}

\renewcommand{\section}[1]{\emph{#1.---}}

\begin{document}

\newcommand{\myTitle}{Theory of reentrant superconductivity in Corbino Josephson junctions}

\newcommand{\Weizmann}{Department of Condensed Matter Physics, Weizmann Institute of Science, Rehovot, Israel 7610001}
\newcommand{\Cornell}{Department of Physics, Cornell University, Ithaca, NY 14853, USA}
\newcommand{\HarvardPhysics}{Department of Physics, Harvard University, Cambridge, Massachusetts 02138, USA}
\newcommand{\HarvardEngineering}{John A. Paulson School of Engineering and Applied Sciences, Harvard University, Cambridge, Massachusetts 02138, USA}
\newcommand{\Columbia}{Department of Applied Physics and Applied Mathematics, Columbia University, New York, New York 10027, USA}
\newcommand{\SKKU}{Department of Physics, Sungkyunkwan University (SKKU), Suwon 16419, Republic of Korea}
\newcommand{\IBS}{Center for 2D Quantum Heterostructures, Institute for Basic Science (IBS), Sungkyunkwan University (SKKU), Suwon 16419, Republic of Korea}

\author{Omri Lesser}
\affiliation{\Cornell}
\affiliation{\Weizmann}

\author{Joon Young Park}
\affiliation{\HarvardPhysics}
\affiliation{\SKKU}
\affiliation{\IBS}

\author{Yuval Ronen}
\affiliation{\Weizmann}

\author{Thomas Werkmeister}
\affiliation{\HarvardEngineering}
\affiliation{\Columbia}

\author{Philip Kim}
\affiliation{\HarvardPhysics}
\affiliation{\HarvardEngineering}

\author{Yuval Oreg}
\affiliation{\Weizmann}
\title{\myTitle}

\begin{abstract}
Josephson junctions made of conventional superconductors display Fraunhofer-like oscillations of the critical current as a function of the threaded magnetic flux. When the superconductors are deposited on the surface of a three-dimensional topological insulator, this pattern is slightly modified due to the presence of chiral Majorana modes. Here we calculate the critical current of a Corbino Josephson junction, where the fluxoid becomes quantized and the superconducting phase has an integer winding.
We discover that circular junctions exhibit similar behavior in both topologically trivial and non-trivial scenarios, while non-circular junctions demonstrate a remarkable distinction.
Using a simple analytical model, we show that these non-circular junctions exhibit reentrant superconductivity with a period related to their number of corners, and numerically we find that this period is halved in the topological case. The period halving may help establish the existence of topological superconductivity in hybrid topological insulator--superconductor junctions.
\end{abstract}
\maketitle

\section{Introduction}Josephson junctions (JJs) are interfaces between two superconductors (SCs) separated by a normal conducting region~\cite{tinkham_introduction_2004}. The normal region allows the formation of Andreev bound states, whose energies $E$ reside inside the superconducting gap and depend on the gauge-invariant phase difference between the SCs,~$\phi$. The supercurrent associated with these states is proportional to $dE/d\phi$, which in a conventional junction is approximately $\propto\sin\left(\phi\right)$. When magnetic flux $\Phi$ is threaded through the junction, supercurrent can flow without dissipation below the critical current $I_c\left(\Phi\right)$, above which the junction becomes resistive. Physically, the phase difference between the SCs adjusts itself to sustain the maximal current, resulting in the celebrated Fraunhofer-like pattern 
\begin{equation}\label{eq:Fraunhofer}
    I_c(\Phi) = I_c(0)\left|\frac{\sin\left(\pi\Phi/\Phi_0\right)}{\pi\Phi/\Phi_0}\right|,
\end{equation}
where $\Phi_0=h/2e$ is the superconducting flux quantum.

Recently, interest in JJs with topological properties has increased~\cite{fu_superconducting_2008,qi_topological_2011,alicea_new_2012,leijnse_introduction_2012,hell_two-dimensional_2017,pientka_topological_2017}. Junctions made of topological superconductors are expected to exhibit unusual periodicity because of the existence of Majorana bound states. An intriguing route to realizing topological JJs is placing conventional SCs in proximity to three-dimensional topological insulators (3DTIs), which have a gapped bulk and metallic, Dirac-like, surface states~\cite{hasan_colloquium_2010,bernevig_topological_2013}. Potter and Fu~\cite{potter_anomalous_2013} have studied the Fraunhofer pattern in a planar JJ on top of a 3DTI surface and found node lifting, i.e., the critical current does not go all the way to zero at integer fluxes. However, the deviation from zero is hard to discern experimentally. 

JJs on 3DTIs have been studied both theoretically~\cite{grosfeld_observing_2011,potter_anomalous_2013,park_detecting_2015,hegde_topological_2020,abboud_signatures_2022,okugawa_vortex_2022} and experimentally~\cite{veldhorst_josephson_2012,williams_unconventional_2012,cho_symmetry_2013,lee_local_2014,kurter_evidence_2015,charpentier_induced_2017,ghatak_anomalous_2018,kayyalha_anomalous_2019,kayyalha_highly_2020,takeshige_experimental_2020}. Several of these studies focused on the Fraunhofer-like pattern of the critical current, many times finding behavior similar to that of conventional JJs. However, in some cases, anomalous characteristics appeared in the Fraunhofer pattern, such as a low value for the edge of the first lobe~\cite{williams_unconventional_2012} or missing lobes~\cite{ghatak_anomalous_2018}. These promising findings encourage further exploration of JJs on 3DTIs, with the hope of accessing the current-phase relation more directly.

In this manuscript, we study the critical current in Corbino JJs, where the geometry is closed (see Fig.~\ref{fig:schematic}) and therefore fluxoid quantization holds~\cite{hadfield_corbino_2003,clem_corbino-geometry_2010}. This geometry has been explored recently in a circular InAs/Al junction~\cite{matsuo_evaluation_2020}, but is significantly less studied than planar geometries~\cite{dominguez_fraunhofer_2022}. In 3DTIs, Corbino JJs enable probing of a \emph{single surface}, provided that the bulk gap is large. We first characterize the behavior of the critical current as a function of the flux in conventional JJs, and find a periodic reentrance of superconductivity which depends on the junction's geometry. We then extend the model of Ref.~\cite{potter_anomalous_2013} to study the topological counterpart of these JJs, and find a \emph{period halving} in the reentrant superconductivity. We establish Corbino JJs as experimentally accessible probes of the current-phase relation and, thus, a potential marker of non-trivial topology.

\section{Corbino Josephson junctions}The simplest way to go from a planar JJ to Corbino geometry is to impose fluxoid quantization. To do this, we take the Fraunhofer pattern Eq.~\eqref{eq:Fraunhofer} and assert that only integer values of $\Phi/\Phi_0\equiv n_{\rm v}$ are allowed. Now $n_{\rm v}$ represents the number of (Josephson) vortices penetrating the normal region between the inner and outer SC. Since $\sin\left(\pi n_{\rm v}\right)=0$ for any nonzero integer $n_{\rm v}$ (similar to the node structure appearing in planar JJs), we find that superconductivity is completely destroyed as soon as any vortices enter the junction.
This finding originates from two assumptions: a conventional current-phase relation, $I\propto\sin\left(\phi\right)$, and linear phase growth along the angular coordinate, $\phi\left(\theta\right)\propto\theta$. These are reasonable assumptions for a \emph{circular} Corbino JJ with a conventional metallic normal region.
We will soon see that if one or both of these assumptions are broken, richer features emerge.

\begin{figure}
    \centering
   \includegraphics[width=\linewidth]{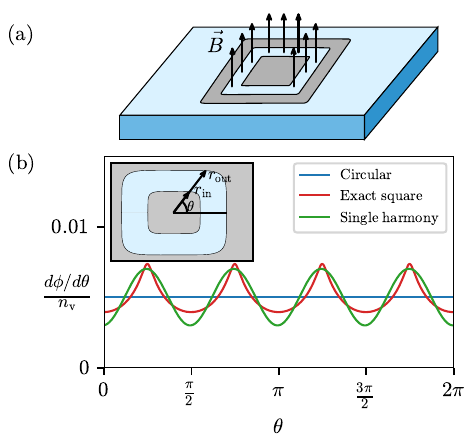}
    \caption{(a)~Illustration of a Corbino Josephson junction. Blue: normal region (taken to be either a standard metal or a three-dimensional topological insulator). Gray: superconducting pads. The out-of-plane magnetic field $\vec{B}$ penetrates the normal region between the superconductors.
    (b)~Phase gradient as a function of the coordinate $\theta$. For a circular junction the gradient is constant, whereas the square junction shows kinks due to the corners. The simplified single-harmony model of Eq.~\eqref{eq:single_harmony} captures this kinks structure. Inset: top view of the junction with the curves $r_{\rm in}\left(\theta\right)$ and $r_{\rm out}\left(\theta\right)$ annotated; these are the inputs to the exact phase evolution in Eq.~\eqref{eq:general_phase_evolution}.
    \label{fig:schematic}}
\end{figure}

Consider a two-dimensional Corbino JJ where the inner and outer boundaries are represented by the polar functions $r_{\rm in}(\theta)$ and $r_{\rm out}(\theta)$, as depicted in Fig.~\ref{fig:schematic}. Assuming the perpendicular magnetic field only penetrates in the normal region, the phase of the inner SC remains constant. The phase difference between the inner and outer SCs $\phi(\theta)$ is proportional to the flux in the section between $\theta$ and some reference point (conveniently taken as $\theta=0$). Therefore, we find that
\begin{equation}\label{eq:general_phase_evolution}
    \frac{d\phi\left(\theta\right)}{d\theta}=\pi n_{\rm v} \frac{r_{\rm out}^{2}\left(\theta\right)-r_{\rm in}^{2}\left(\theta\right)}{S_{\rm tot}},
\end{equation}
where $n_{\rm v} = \Phi/\Phi_{0}$ is the number of vortices and $S_{\rm tot}=\frac{1}{2}\int_{0}^{2\pi}d\theta\left[r_{\rm out}^{2}\left(\theta\right)-r_{\rm in}^{2}\left(\theta\right)\right]$ is the total area of the normal region. In the special case of a circular junction, i.e., constant $r_{\rm in}$ and $r_{\rm out}$, we obtain a constant phase gradient $d\phi/d\theta=n_{\rm v}$. In Fig.~\figref{fig:schematic}{b} we show the phase evolution for circular and square junctions.

\section{Reentrant superconductivity in conventional non-circular junctions}To find the critical current of a Corbino JJ we assume a conventional current-phase relation and maximize the Josephson current,
\begin{equation}
    I_{c}=I_{0}\max_{\phi_{0}}\int_{0}^{2\pi}d\theta\sin\left[\phi\left(\theta\right)+\phi_{0}\right],
\end{equation}
where $I_0$ is a constant and $\phi_0$ is a global phase offset between the inner and outer superconductors. We emphasize that the junction is fully described by its phase evolution $\phi\left(\theta\right)$, which is determined from the geometry and the number of flux quanta by Eq.~\eqref{eq:general_phase_evolution}.
As a simple model for a rectangular Corbino JJ, we use the polar equation $r_{\rm in,out}\left(\theta\right)=R_{\rm in,out}\left( \cos^{2n}\theta + \sin^{2n}\theta \right)^{-1/2n}$. For $n=1$ we obtain a circle, and for $n\to\infty$ we get a perfect square. Integer values of $n$ interpolate between the two limiting cases.
Plugging these polar forms in Eq.~\eqref{eq:general_phase_evolution} for the phase evolution and numerically solving the integral, we find that the critical current is zero unless $n_{\rm v}$ is a multiple of four.

To gain analytical understanding of the result, consider the following simple model of the phase evolution in a junction with $n_{\rm c}$ corners:
\begin{equation}\label{eq:single_harmony}
    \phi(\theta) = n_{\rm v}\theta + a\sin(n_{\rm c}\theta),
\end{equation}
where $a$ is a constant. For $a=0$ we get a constant phase gradient, corresponding to a circular junction; $a\neq0$ is the perturbation due to the $n_{\rm c}$ corners (for a general shape, $n_{\rm c}$ represents half the number of zeros of $d^2\phi/d\theta^2$). Notice that $\phi\left(0\right)=0$ and $\phi\left(2\pi\right)=2\pi n_{\rm v}$ such that the perturbation leaves the endpoints intact. We assume $|a|<n_{\rm v}/n_{\rm c}$ such that $\phi(\theta)$ is monotonically increasing.

The critical current in this model for a conventional junction is given by
\begin{equation}\label{eq:I_c_Bessel}
    \begin{aligned}
    I_c &= I_0 \max_{\phi_{0}}\int_{0}^{2\pi}d\theta\sin\left[n_{\rm v}\theta+a\sin\left(n_{\rm c}\theta\right)+\phi_{0}\right]\\
    &= I_0 \sum_{m=0}^{\infty} (-1)^{m} J_{m}(a) \int_{0}^{2\pi}d\theta \sin\left[ \left(n_{\rm v}-mn_{\rm c}\right)\theta \right],
\end{aligned}
\end{equation}
where and $J_{m}$ is the $m^{\rm th}$ Bessel function of the first kind. From Eq.~\eqref{eq:I_c_Bessel} we see that the critical current can only be nonzero if $n_{\rm v}=mn_{\rm c}$ for some integer $m$, i.e., if the number of vortices is an integer multiple of the number of corners (and then $\phi_0$ takes the value $\pi/2$). This is exactly the result we obtained numerically when taking into account the exact phase evolution. This simple model, whose mathematical analysis is very similar to that of the Shapiro steps~\cite{shapiro_josephson_1963,tinkham_introduction_2004}, therefore contains the relevant physics for understanding the behavior of the critical current [see Fig.~\figref{fig:schematic}{b} for a comparison of the phase evolution in this model with the exact one].
Intuitively, the critical current vanishes due to the cancellation of local supercurrents, except at special values of the flux at which the supercurrent distribution at the corners counteracts this effect.

\begin{figure}
    \centering
   \includegraphics[width=\linewidth]{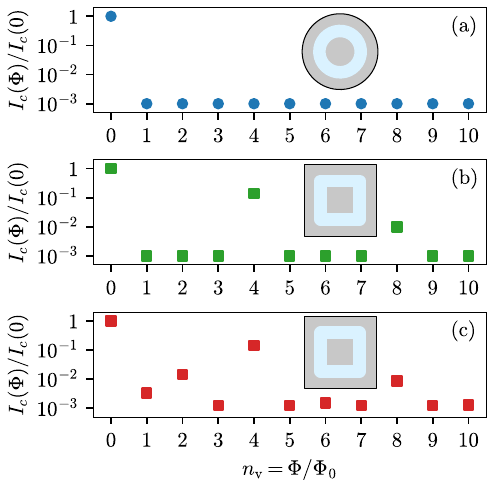}
    \caption{Critical current as a function of the number of vortices $n_{\rm v}$ for (a)~a circular Corbino Josephson junction, (b)~a conventional square junction, and (c)~a topological square junction.
    The circular junction behaves identically for the conventional and topological cases: any $n_{\rm v}>0$ completely destroys superconductivity.
    In the square junction, the critical current is nonzero when $n_{\rm v}$ is a multiple of four, with the topological junction having nonzero $I_c$ also when $n_{\rm v}$ is a multiple of two.
    All values are normalized by their respective zero-flux critical current $I_c\left(0\right)$.
    For clarity of the logarithmic scale, we cut off the signals from below at $10^{-3}$.
    \label{fig:I_c_all}}
\end{figure}

\section{Topological Corbino junctions}Having established the behavior of circular and non-circular conventional JJs, we turn our attention to Corbino JJs where the normal part is the surface of a 3DTI. We will construct a model for the edge states along the lines of Ref.~\cite{potter_anomalous_2013}, and then plug in the general phase evolution from Eq.~\eqref{eq:general_phase_evolution}. By numerically diagonalizing the Hamiltonian, we will find the critical current and compare it to the non-topological case.

The low-energy physics of a 3DTI surface state coupled to two SCs is described by two counter-propagating one-dimensional Majorana modes~\cite{fu_superconducting_2008,potter_anomalous_2013} $\gamma_{\rm in}$, $\gamma_{\rm out}$. The effective Hamiltonian reads 
\begin{equation}\label{eq:H_eff_Majoranas}
\begin{aligned}H_{\text{eff}}=\int_{0}^{2\pi}d\theta & \left[iv\left(\gamma_{\text{in}}\left(\theta\right)\partial_{\theta}\gamma_{\text{in}}\left(\theta\right)-\gamma_{\text{out}}\left(\theta\right)\partial_{\theta}\gamma_{\text{out}}\left(\theta\right)\right)\right.\\
 & \left.+i\Delta\cos\left(\frac{\phi\left(\theta\right)}{2}\right)\gamma_{\text{in}}\left(\theta\right)\gamma_{\text{out}}\left(\theta\right)\right].
\end{aligned}
\end{equation}
Here $v$ is the velocity of the Majorana modes (we assume it is equal for the inner and outer modes) and $\Delta$ is the SC pair potential. The $\Delta$ term locally couples the two modes according to the local phase difference $\phi\left(\theta\right)$. This low-energy description is valid in the narrow junction limit where the width of the normal region $W$ is shorter than the coherence length $\xi$. 

\begin{figure}
    \centering
   \includegraphics[width=\linewidth]{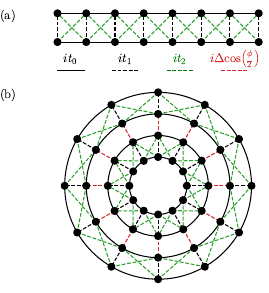}
    \caption{(a)~Majorana ladder corresponding to the variant of the Grover--Sheng--Vishwanath Hamiltonian~\cite{grover_emergent_2014,li_coupled_2020} we adopt, Eq.~\eqref{eq:H_ladder}. When $t_1=2t_2$ two Majorana modes of opposite chirality are localized each on a different chain.
    (b)~Two such ladders arranged in a closed geometry to resemble the inner and outer modes of a Corbino JJ on a 3DTI. The Josephson coupling between the inner and outer modes (dashed red lines) depends on the local phase difference $\phi\left(\theta\right)$.
    \label{fig:ladders}}
\end{figure}

For general $\phi\left(\theta\right)$, the Hamiltonian Eq.~\eqref{eq:H_eff_Majoranas} can be diagonalized numerically by discretizing it and solving the tight-binding problem. However, this requires some care, since chiral Majorana modes cannot be simulated by themselves, due to the Nielsen--Ninomiya theorem~\cite{nielsen_absence_1981,nielsen_absence_1981-1}. 
We get around this problem using a variant of the Grover--Sheng--Vishwanath model~\cite{grover_emergent_2014,li_coupled_2020}, which is a two-leg ladder of Majorana fermions with intra-chain hopping $t_0$ and two types of inter-chain hoppings $t_1$, $t_2$, see Fig.~\figref{fig:ladders}{a}. The model is described by the Hamiltonian 
\begin{equation}
\begin{aligned}\label{eq:H_ladder}
    H_{\rm ladder} = \sum_{j=1}^{N}&\left[it_0\left(\alpha_{j}\alpha_{j+1}-\beta_{j}\beta_{j+1}\right) \right. \\ 
    &\left. +it_{1}\beta_{j}\alpha_{j}+it_{2}\left(\alpha_{j}\beta_{j+1}-\beta_{j}\alpha_{j+1}\right)\right],
\end{aligned}
\end{equation}
where $\alpha$, $\beta$ are Majorana operators at the two chains and $N$ is the number of sites per chain (we identify site $N+1$ with site $1$ to get periodic boundary conditions).
At $t_{1}=2t_{2}$ the spectrum becomes gapless, with two modes of opposite chirality residing in the two chains that comprise the ladder~\cite{li_coupled_2020}. It is now possible to add a term coupling these modes along the lines of the continuum Hamiltonian~\eqref{eq:H_eff_Majoranas}.

However, the presence of vortices brings about another complication: in a closed geometry, a Majorana mode encircling $n_{\rm v}$ vortices will have periodic boundary conditions for even $n_{\rm v}$ and anti-periodic boundary conditions for odd $n_{\rm v}$. Under our assumptions, the inner mode $\gamma_{\rm in}$ does not encircle any flux, so it always has periodic boundary conditions, but for the outer mode $\gamma_{\rm out}$, both options are possible. These mixed boundary conditions cannot be realized in a single ladder as in Eq.~\eqref{eq:H_ladder}. We, therefore, use two copies of $H_{\rm ladder}$, the ``inner" one having periodic boundary conditions and the ``outer" one having either periodic or anti-periodic boundary conditions depending on $n_{\rm v}$ (the anti-periodic version is realized by flipping the signs of $t_0$ and $t_2$ at the last bond), see Fig.~\figref{fig:ladders}{b}. Finally, we introduce the Josephson coupling 
\begin{equation}\label{eq:H_Delta}
    H_{\Delta}=\sum_{j=1}^{N}i\Delta\cos\left[\frac{\phi_0 + \phi\left(\frac{2\pi}{N}j\right)}{2}\right]\alpha_{j}^{\rm (in)}\beta_{j}^{\rm (out)}.
\end{equation}
This completes the tight-binding discretization of the continuum Hamiltonian~\eqref{eq:H_eff_Majoranas}. We note that the modes $\beta^{\rm (in)}$, $\alpha^{\rm (out)}$ are unaffected by $H_{\Delta}$ and remain gapless --- they are redundancies of the model.

With the model set, we are ready to calculate the critical current for different Corbino geometries, as determined by the phase evolution $\phi\left(\theta\right)$. To do so, we diagonalize the Hamiltonian numerically and find the eigen-energies $\left\{ E_{\ell} \right\}$. The Josephson current as a function of the global phase offset $\phi_0$ is then given by~\cite{tinkham_introduction_2004,setiawan_topological_2019}
\begin{equation}
    I\left(\phi_0\right) = -\frac{4e}{\hbar}\sum_{\ell} \tanh\left(\frac{E_{\ell}}{2k_{\rm B}T}\right) \frac{dE_{\ell}}{d\phi_0},
\end{equation}
where $T$ is the temperature (below we will work in the $T=0$ limit). Finally, the critical current is given by $I_{c}=\max_{\phi_{0}} I\left(\phi_0\right)$. 
We cut off the summation at $\left|E_{\ell}\right|<5\Delta$ to eliminate artificial contributions originating from the redundancies in the lattice model.
The simulations are performed with 400 sites, and the tight-binding parameters are $t_0=1$, $t_1=0.6$, $t_2=0.3$, $\Delta=0.2$.

For a circular topological junction, we find similar behavior to the non-topological case: the critical current is only finite at $n_{\rm v}=0$, and there is no reentrant superconductivity. However, for non-circular junctions, we find a qualitative difference. Whereas in the non-topological case superconductivity reenters for $n_{\rm v}=mn_{\rm c}$ with $n_{\rm c}$ being the number of corners, in the topological case the condition is $n_{\rm v}=\frac{m}{2}n_{\rm c}$, i.e., the reentrance period is \emph{halved}. We demonstrate this for a square junction ($n_{\rm c}=4$) in Fig.~\ref{fig:I_c_all}: the topological junction shows $I_{c}\neq0$ for all even $n_{\rm v}$ where the non-topological junction only does so for multiples of four.
Notice that this effect only occurs when $n_{\rm c}$ is \emph{even}; for odd $n_{\rm c}$ the topological and conventional junctions behave qualitatively the same.
As we show explicitly in the Supplemental Material~\cite{SupplementalMaterial}, such period halving is consistent with a $\sin\left(2\phi\right)$ term in the current-phase relation.
We further find that upon breaking inversion symmetry, the Corbino junction exhibits a Josephson diode effect, with even-odd polarity switches as a function of $n_{\rm v}$. This behavior is unique to junctions whose low-energy physics is described by Majorana modes, and it is consistent with an upcoming experimental manuscript~\cite{park_2025}.

\section{Outlook}We have shown that the critical current in Corbino JJs conveys valuable information about the nature of the junction. In particular, non-circular junctions exhibit periodic reentrance of superconductivity as a function of the applied flux, with the periodicity depending on the geometry of the junction. In a non-circular Corbino JJ on the surface of a 3DTI, this periodicity is halved.

Several limitations of our modeling are noteworthy. First, we assumed that the low-energy physics of the 3DTI JJ is completely described in terms of the surface states, without mixing with the bulk states. This is a good assumption only as long as the bulk gap of the 3DTI is large, and in particular much larger than the superconducting pair potential; this is indeed the case for state of the art 3DTI crystals~\cite{kushwaha_sn-doped_2016}. We have also assumed that the flux only enters the normal part of the junction, which is valid if the Meissner effect is sufficiently strong.

From a practical, experimental point of view, additional considerations we neglected may play a role. If the superconductor is thin, then flux trapping is possible, and one must take into account the vortex dynamics in the junction. This complication may be relieved if vortices have naturally favorable points in space to enter the junction through, such as defects or regions with a weaker proximity effect. In addition, we note that the period halving effect is most visible in the first half-harmonic $n_{\rm v}=n_{\rm c}/2$, since the envelope of the critical current decays roughly as $1/n_{\rm v}$. One would therefore have to rely mostly on this first half-harmonic when claiming topological behavior in such junctions. Even then, we acknowledge that alternative mechanisms might lead to period halving; although we could not identify such alternatives, we cannot rule out their existence. Finally, an important experimental complication is the existence of contacts to the inner and outer SCs, which may induce disorder and alter the mobility of the surface state. This effect has been neglected in our study, in the hope that the as long as the bulk gap remains open, the surface states maintain their robustness and the qualitative picture does not change.

Despite these limitations, non-circular Corbino JJs bring valuable new information to the table. The main benefit of our approach is that such junctions provide a glance at the underlying current-phase relation. The stark difference between topological and conventional junctions in these geometries could help advance the search for conclusive evidence of topological superconductivity.

\section{Acknowledgment}We thank Alexander Shnirman for useful discussions.
This work was supported by the European Union's Horizon 2020 research and innovation programme (Grant Agreement LEGOTOP No. 788715), the DFG (CRC/Transregio 183, EI 519/7-1), the Israel Science
Foundation ISF 1914/24, and ISF-Mafat 2478/24.
J.Y.P. acknowledges support from the Institute for Basic Science (IBS-R036-D1) and the National Research Foundation of Korea (NRF) grant funded by the Korean government (MSIT) (No. RS-2021-NR060087).
T.W., and P.K. acknowledge support from the ONR (N00014-24-1-2081). T.W. acknowledges support from the Simons Foundation Society of Fellows (SFI-MPS-SFJ-00011748).
Y.R. acknowledges support from the European Research Council Starting Investigator Grant No. 101163917; the Minerva Foundation with funding from the Federal German Ministry for Education and Research; and the Israel Science Foundation under Grant Nos. 380/25 and 425/25.

\bibliography{library}

\begin{thebibliography}{38}%
\makeatletter
\providecommand \@ifxundefined [1]{%
 \@ifx{#1\undefined}
}%
\providecommand \@ifnum [1]{%
 \ifnum #1\expandafter \@firstoftwo
 \else \expandafter \@secondoftwo
 \fi
}%
\providecommand \@ifx [1]{%
 \ifx #1\expandafter \@firstoftwo
 \else \expandafter \@secondoftwo
 \fi
}%
\providecommand \natexlab [1]{#1}%
\providecommand \enquote  [1]{``#1''}%
\providecommand \bibnamefont  [1]{#1}%
\providecommand \bibfnamefont [1]{#1}%
\providecommand \citenamefont [1]{#1}%
\providecommand \href@noop [0]{\@secondoftwo}%
\providecommand \href [0]{\begingroup \@sanitize@url \@href}%
\providecommand \@href[1]{\@@startlink{#1}\@@href}%
\providecommand \@@href[1]{\endgroup#1\@@endlink}%
\providecommand \@sanitize@url [0]{\catcode `\\12\catcode `\$12\catcode `\&12\catcode `\#12\catcode `\^12\catcode `\_12\catcode `\%12\relax}%
\providecommand \@@startlink[1]{}%
\providecommand \@@endlink[0]{}%
\providecommand \url  [0]{\begingroup\@sanitize@url \@url }%
\providecommand \@url [1]{\endgroup\@href {#1}{\urlprefix }}%
\providecommand \urlprefix  [0]{URL }%
\providecommand \Eprint [0]{\href }%
\providecommand \doibase [0]{https://doi.org/}%
\providecommand \selectlanguage [0]{\@gobble}%
\providecommand \bibinfo  [0]{\@secondoftwo}%
\providecommand \bibfield  [0]{\@secondoftwo}%
\providecommand \translation [1]{[#1]}%
\providecommand \BibitemOpen [0]{}%
\providecommand \bibitemStop [0]{}%
\providecommand \bibitemNoStop [0]{.\EOS\space}%
\providecommand \EOS [0]{\spacefactor3000\relax}%
\providecommand \BibitemShut  [1]{\csname bibitem#1\endcsname}%
\let\auto@bib@innerbib\@empty
\bibitem [{\citenamefont {Tinkham}(1996)}]{tinkham_introduction_2004}%
  \BibitemOpen
  \bibfield  {author} {\bibinfo {author} {\bibfnamefont {M.}~\bibnamefont {Tinkham}},\ }\href@noop {} {\emph {\bibinfo {title} {Introduction to Superconductivity}}},\ \bibinfo {edition} {2nd}\ ed.,\ International Series in Pure and Applied Physics\ (\bibinfo  {publisher} {{McGraw-Hill}},\ \bibinfo {address} {{New York}},\ \bibinfo {year} {1996})\BibitemShut {NoStop}%
\bibitem [{\citenamefont {Fu}\ and\ \citenamefont {Kane}(2008)}]{fu_superconducting_2008}%
  \BibitemOpen
  \bibfield  {author} {\bibinfo {author} {\bibfnamefont {L.}~\bibnamefont {Fu}}\ and\ \bibinfo {author} {\bibfnamefont {C.~L.}\ \bibnamefont {Kane}},\ }\href {https://doi.org/10.1103/PhysRevLett.100.096407} {\bibfield  {journal} {\bibinfo  {journal} {Physical Review Letters}\ }\textbf {\bibinfo {volume} {100}},\ \bibinfo {pages} {096407} (\bibinfo {year} {2008})}\BibitemShut {NoStop}%
\bibitem [{\citenamefont {Qi}\ and\ \citenamefont {Zhang}(2011)}]{qi_topological_2011}%
  \BibitemOpen
  \bibfield  {author} {\bibinfo {author} {\bibfnamefont {X.-L.}\ \bibnamefont {Qi}}\ and\ \bibinfo {author} {\bibfnamefont {S.-C.}\ \bibnamefont {Zhang}},\ }\href {https://doi.org/10.1103/RevModPhys.83.1057} {\bibfield  {journal} {\bibinfo  {journal} {Reviews of Modern Physics}\ }\textbf {\bibinfo {volume} {83}},\ \bibinfo {pages} {1057} (\bibinfo {year} {2011})}\BibitemShut {NoStop}%
\bibitem [{\citenamefont {Alicea}(2012)}]{alicea_new_2012}%
  \BibitemOpen
  \bibfield  {author} {\bibinfo {author} {\bibfnamefont {J.}~\bibnamefont {Alicea}},\ }\href {https://doi.org/10.1088/0034-4885/75/7/076501} {\bibfield  {journal} {\bibinfo  {journal} {Reports on Progress in Physics}\ }\textbf {\bibinfo {volume} {75}},\ \bibinfo {pages} {076501} (\bibinfo {year} {2012})}\BibitemShut {NoStop}%
\bibitem [{\citenamefont {Leijnse}\ and\ \citenamefont {Flensberg}(2012)}]{leijnse_introduction_2012}%
  \BibitemOpen
  \bibfield  {author} {\bibinfo {author} {\bibfnamefont {M.}~\bibnamefont {Leijnse}}\ and\ \bibinfo {author} {\bibfnamefont {K.}~\bibnamefont {Flensberg}},\ }\href {https://doi.org/10.1088/0268-1242/27/12/124003} {\bibfield  {journal} {\bibinfo  {journal} {Semiconductor Science and Technology}\ }\textbf {\bibinfo {volume} {27}},\ \bibinfo {pages} {124003} (\bibinfo {year} {2012})}\BibitemShut {NoStop}%
\bibitem [{\citenamefont {Hell}\ \emph {et~al.}(2017)\citenamefont {Hell}, \citenamefont {Leijnse},\ and\ \citenamefont {Flensberg}}]{hell_two-dimensional_2017}%
  \BibitemOpen
  \bibfield  {author} {\bibinfo {author} {\bibfnamefont {M.}~\bibnamefont {Hell}}, \bibinfo {author} {\bibfnamefont {M.}~\bibnamefont {Leijnse}},\ and\ \bibinfo {author} {\bibfnamefont {K.}~\bibnamefont {Flensberg}},\ }\href {https://doi.org/10.1103/PhysRevLett.118.107701} {\bibfield  {journal} {\bibinfo  {journal} {Physical Review Letters}\ }\textbf {\bibinfo {volume} {118}},\ \bibinfo {pages} {107701} (\bibinfo {year} {2017})}\BibitemShut {NoStop}%
\bibitem [{\citenamefont {Pientka}\ \emph {et~al.}(2017)\citenamefont {Pientka}, \citenamefont {Keselman}, \citenamefont {Berg}, \citenamefont {Yacoby}, \citenamefont {Stern},\ and\ \citenamefont {Halperin}}]{pientka_topological_2017}%
  \BibitemOpen
  \bibfield  {author} {\bibinfo {author} {\bibfnamefont {F.}~\bibnamefont {Pientka}}, \bibinfo {author} {\bibfnamefont {A.}~\bibnamefont {Keselman}}, \bibinfo {author} {\bibfnamefont {E.}~\bibnamefont {Berg}}, \bibinfo {author} {\bibfnamefont {A.}~\bibnamefont {Yacoby}}, \bibinfo {author} {\bibfnamefont {A.}~\bibnamefont {Stern}},\ and\ \bibinfo {author} {\bibfnamefont {B.~I.}\ \bibnamefont {Halperin}},\ }\href {https://doi.org/10.1103/PhysRevX.7.021032} {\bibfield  {journal} {\bibinfo  {journal} {Physical Review X}\ }\textbf {\bibinfo {volume} {7}},\ \bibinfo {pages} {021032} (\bibinfo {year} {2017})}\BibitemShut {NoStop}%
\bibitem [{\citenamefont {Hasan}\ and\ \citenamefont {Kane}(2010)}]{hasan_colloquium_2010}%
  \BibitemOpen
  \bibfield  {author} {\bibinfo {author} {\bibfnamefont {M.~Z.}\ \bibnamefont {Hasan}}\ and\ \bibinfo {author} {\bibfnamefont {C.~L.}\ \bibnamefont {Kane}},\ }\href {https://doi.org/10.1103/RevModPhys.82.3045} {\bibfield  {journal} {\bibinfo  {journal} {Reviews of Modern Physics}\ }\textbf {\bibinfo {volume} {82}},\ \bibinfo {pages} {3045} (\bibinfo {year} {2010})}\BibitemShut {NoStop}%
\bibitem [{\citenamefont {Bernevig}\ and\ \citenamefont {Hughes}(2013)}]{bernevig_topological_2013}%
  \BibitemOpen
  \bibfield  {author} {\bibinfo {author} {\bibfnamefont {B.~A.}\ \bibnamefont {Bernevig}}\ and\ \bibinfo {author} {\bibfnamefont {T.~L.}\ \bibnamefont {Hughes}},\ }\href@noop {} {\emph {\bibinfo {title} {Topological Insulators and Topological Superconductors}}}\ (\bibinfo  {publisher} {{Princeton university press}},\ \bibinfo {year} {2013})\BibitemShut {NoStop}%
\bibitem [{\citenamefont {Potter}\ and\ \citenamefont {Fu}(2013)}]{potter_anomalous_2013}%
  \BibitemOpen
  \bibfield  {author} {\bibinfo {author} {\bibfnamefont {A.~C.}\ \bibnamefont {Potter}}\ and\ \bibinfo {author} {\bibfnamefont {L.}~\bibnamefont {Fu}},\ }\href {https://doi.org/10.1103/PhysRevB.88.121109} {\bibfield  {journal} {\bibinfo  {journal} {Physical Review B}\ }\textbf {\bibinfo {volume} {88}},\ \bibinfo {pages} {121109} (\bibinfo {year} {2013})}\BibitemShut {NoStop}%
\bibitem [{\citenamefont {Grosfeld}\ and\ \citenamefont {Stern}(2011)}]{grosfeld_observing_2011}%
  \BibitemOpen
  \bibfield  {author} {\bibinfo {author} {\bibfnamefont {E.}~\bibnamefont {Grosfeld}}\ and\ \bibinfo {author} {\bibfnamefont {A.}~\bibnamefont {Stern}},\ }\href {https://doi.org/10.1073/pnas.1101469108} {\bibfield  {journal} {\bibinfo  {journal} {Proceedings of the National Academy of Sciences}\ }\textbf {\bibinfo {volume} {108}},\ \bibinfo {pages} {11810} (\bibinfo {year} {2011})}\BibitemShut {NoStop}%
\bibitem [{\citenamefont {Park}\ and\ \citenamefont {Recher}(2015)}]{park_detecting_2015}%
  \BibitemOpen
  \bibfield  {author} {\bibinfo {author} {\bibfnamefont {S.}~\bibnamefont {Park}}\ and\ \bibinfo {author} {\bibfnamefont {P.}~\bibnamefont {Recher}},\ }\href {https://doi.org/10.1103/PhysRevLett.115.246403} {\bibfield  {journal} {\bibinfo  {journal} {Physical Review Letters}\ }\textbf {\bibinfo {volume} {115}},\ \bibinfo {pages} {246403} (\bibinfo {year} {2015})}\BibitemShut {NoStop}%
\bibitem [{\citenamefont {Hegde}\ \emph {et~al.}(2020)\citenamefont {Hegde}, \citenamefont {Yue}, \citenamefont {Wang}, \citenamefont {Huemiller}, \citenamefont {Van~Harlingen},\ and\ \citenamefont {Vishveshwara}}]{hegde_topological_2020}%
  \BibitemOpen
  \bibfield  {author} {\bibinfo {author} {\bibfnamefont {S.~S.}\ \bibnamefont {Hegde}}, \bibinfo {author} {\bibfnamefont {G.}~\bibnamefont {Yue}}, \bibinfo {author} {\bibfnamefont {Y.}~\bibnamefont {Wang}}, \bibinfo {author} {\bibfnamefont {E.}~\bibnamefont {Huemiller}}, \bibinfo {author} {\bibfnamefont {D.~J.}\ \bibnamefont {Van~Harlingen}},\ and\ \bibinfo {author} {\bibfnamefont {S.}~\bibnamefont {Vishveshwara}},\ }\href {https://doi.org/10.1016/j.aop.2020.168326} {\bibfield  {journal} {\bibinfo  {journal} {Annals of Physics}\ }\textbf {\bibinfo {volume} {423}},\ \bibinfo {pages} {168326} (\bibinfo {year} {2020})}\BibitemShut {NoStop}%
\bibitem [{\citenamefont {Abboud}\ \emph {et~al.}(2022)\citenamefont {Abboud}, \citenamefont {Subramanyan}, \citenamefont {Sun}, \citenamefont {Yue}, \citenamefont {Van~Harlingen},\ and\ \citenamefont {Vishveshwara}}]{abboud_signatures_2022}%
  \BibitemOpen
  \bibfield  {author} {\bibinfo {author} {\bibfnamefont {N.}~\bibnamefont {Abboud}}, \bibinfo {author} {\bibfnamefont {V.}~\bibnamefont {Subramanyan}}, \bibinfo {author} {\bibfnamefont {X.-Q.}\ \bibnamefont {Sun}}, \bibinfo {author} {\bibfnamefont {G.}~\bibnamefont {Yue}}, \bibinfo {author} {\bibfnamefont {D.}~\bibnamefont {Van~Harlingen}},\ and\ \bibinfo {author} {\bibfnamefont {S.}~\bibnamefont {Vishveshwara}},\ }\href {https://doi.org/10.1103/PhysRevB.105.214521} {\bibfield  {journal} {\bibinfo  {journal} {Physical Review B}\ }\textbf {\bibinfo {volume} {105}},\ \bibinfo {pages} {214521} (\bibinfo {year} {2022})}\BibitemShut {NoStop}%
\bibitem [{\citenamefont {Okugawa}\ \emph {et~al.}(2022)\citenamefont {Okugawa}, \citenamefont {Park}, \citenamefont {Recher},\ and\ \citenamefont {Kennes}}]{okugawa_vortex_2022}%
  \BibitemOpen
  \bibfield  {author} {\bibinfo {author} {\bibfnamefont {T.}~\bibnamefont {Okugawa}}, \bibinfo {author} {\bibfnamefont {S.}~\bibnamefont {Park}}, \bibinfo {author} {\bibfnamefont {P.}~\bibnamefont {Recher}},\ and\ \bibinfo {author} {\bibfnamefont {D.~M.}\ \bibnamefont {Kennes}},\ }\href {https://doi.org/10.1103/PhysRevB.106.024501} {\bibfield  {journal} {\bibinfo  {journal} {Physical Review B}\ }\textbf {\bibinfo {volume} {106}},\ \bibinfo {pages} {024501} (\bibinfo {year} {2022})}\BibitemShut {NoStop}%
\bibitem [{\citenamefont {Veldhorst}\ \emph {et~al.}(2012)\citenamefont {Veldhorst}, \citenamefont {Snelder}, \citenamefont {Hoek}, \citenamefont {Gang}, \citenamefont {Guduru}, \citenamefont {Wang}, \citenamefont {Zeitler}, \citenamefont {{van der Wiel}}, \citenamefont {Golubov}, \citenamefont {Hilgenkamp},\ and\ \citenamefont {Brinkman}}]{veldhorst_josephson_2012}%
  \BibitemOpen
  \bibfield  {author} {\bibinfo {author} {\bibfnamefont {M.}~\bibnamefont {Veldhorst}}, \bibinfo {author} {\bibfnamefont {M.}~\bibnamefont {Snelder}}, \bibinfo {author} {\bibfnamefont {M.}~\bibnamefont {Hoek}}, \bibinfo {author} {\bibfnamefont {T.}~\bibnamefont {Gang}}, \bibinfo {author} {\bibfnamefont {V.~K.}\ \bibnamefont {Guduru}}, \bibinfo {author} {\bibfnamefont {X.~L.}\ \bibnamefont {Wang}}, \bibinfo {author} {\bibfnamefont {U.}~\bibnamefont {Zeitler}}, \bibinfo {author} {\bibfnamefont {W.~G.}\ \bibnamefont {{van der Wiel}}}, \bibinfo {author} {\bibfnamefont {A.~A.}\ \bibnamefont {Golubov}}, \bibinfo {author} {\bibfnamefont {H.}~\bibnamefont {Hilgenkamp}},\ and\ \bibinfo {author} {\bibfnamefont {A.}~\bibnamefont {Brinkman}},\ }\href {https://doi.org/10.1038/nmat3255} {\bibfield  {journal} {\bibinfo  {journal} {Nature Materials}\ }\textbf {\bibinfo {volume} {11}},\ \bibinfo {pages} {417} (\bibinfo {year} {2012})}\BibitemShut {NoStop}%
\bibitem [{\citenamefont {Williams}\ \emph {et~al.}(2012)\citenamefont {Williams}, \citenamefont {Bestwick}, \citenamefont {Gallagher}, \citenamefont {Hong}, \citenamefont {Cui}, \citenamefont {Bleich}, \citenamefont {Analytis}, \citenamefont {Fisher},\ and\ \citenamefont {{Goldhaber-Gordon}}}]{williams_unconventional_2012}%
  \BibitemOpen
  \bibfield  {author} {\bibinfo {author} {\bibfnamefont {J.~R.}\ \bibnamefont {Williams}}, \bibinfo {author} {\bibfnamefont {A.~J.}\ \bibnamefont {Bestwick}}, \bibinfo {author} {\bibfnamefont {P.}~\bibnamefont {Gallagher}}, \bibinfo {author} {\bibfnamefont {S.~S.}\ \bibnamefont {Hong}}, \bibinfo {author} {\bibfnamefont {Y.}~\bibnamefont {Cui}}, \bibinfo {author} {\bibfnamefont {A.~S.}\ \bibnamefont {Bleich}}, \bibinfo {author} {\bibfnamefont {J.~G.}\ \bibnamefont {Analytis}}, \bibinfo {author} {\bibfnamefont {I.~R.}\ \bibnamefont {Fisher}},\ and\ \bibinfo {author} {\bibfnamefont {D.}~\bibnamefont {{Goldhaber-Gordon}}},\ }\href {https://doi.org/10.1103/PhysRevLett.109.056803} {\bibfield  {journal} {\bibinfo  {journal} {Physical Review Letters}\ }\textbf {\bibinfo {volume} {109}},\ \bibinfo {pages} {056803} (\bibinfo {year} {2012})}\BibitemShut {NoStop}%
\bibitem [{\citenamefont {Cho}\ \emph {et~al.}(2013)\citenamefont {Cho}, \citenamefont {Dellabetta}, \citenamefont {Yang}, \citenamefont {Schneeloch}, \citenamefont {Xu}, \citenamefont {Valla}, \citenamefont {Gu}, \citenamefont {Gilbert},\ and\ \citenamefont {Mason}}]{cho_symmetry_2013}%
  \BibitemOpen
  \bibfield  {author} {\bibinfo {author} {\bibfnamefont {S.}~\bibnamefont {Cho}}, \bibinfo {author} {\bibfnamefont {B.}~\bibnamefont {Dellabetta}}, \bibinfo {author} {\bibfnamefont {A.}~\bibnamefont {Yang}}, \bibinfo {author} {\bibfnamefont {J.}~\bibnamefont {Schneeloch}}, \bibinfo {author} {\bibfnamefont {Z.}~\bibnamefont {Xu}}, \bibinfo {author} {\bibfnamefont {T.}~\bibnamefont {Valla}}, \bibinfo {author} {\bibfnamefont {G.}~\bibnamefont {Gu}}, \bibinfo {author} {\bibfnamefont {M.~J.}\ \bibnamefont {Gilbert}},\ and\ \bibinfo {author} {\bibfnamefont {N.}~\bibnamefont {Mason}},\ }\href {https://doi.org/10.1038/ncomms2701} {\bibfield  {journal} {\bibinfo  {journal} {Nature Communications}\ }\textbf {\bibinfo {volume} {4}},\ \bibinfo {pages} {1689} (\bibinfo {year} {2013})}\BibitemShut {NoStop}%
\bibitem [{\citenamefont {Lee}\ \emph {et~al.}(2014)\citenamefont {Lee}, \citenamefont {Lee}, \citenamefont {Park}, \citenamefont {Lee}, \citenamefont {Nam}, \citenamefont {Shin}, \citenamefont {Kim},\ and\ \citenamefont {Lee}}]{lee_local_2014}%
  \BibitemOpen
  \bibfield  {author} {\bibinfo {author} {\bibfnamefont {J.~H.}\ \bibnamefont {Lee}}, \bibinfo {author} {\bibfnamefont {G.-H.}\ \bibnamefont {Lee}}, \bibinfo {author} {\bibfnamefont {J.}~\bibnamefont {Park}}, \bibinfo {author} {\bibfnamefont {J.}~\bibnamefont {Lee}}, \bibinfo {author} {\bibfnamefont {S.-G.}\ \bibnamefont {Nam}}, \bibinfo {author} {\bibfnamefont {Y.-S.}\ \bibnamefont {Shin}}, \bibinfo {author} {\bibfnamefont {J.~S.}\ \bibnamefont {Kim}},\ and\ \bibinfo {author} {\bibfnamefont {H.-J.}\ \bibnamefont {Lee}},\ }\href {https://doi.org/10.1021/nl501481b} {\bibfield  {journal} {\bibinfo  {journal} {Nano Letters}\ }\textbf {\bibinfo {volume} {14}},\ \bibinfo {pages} {5029} (\bibinfo {year} {2014})}\BibitemShut {NoStop}%
\bibitem [{\citenamefont {Kurter}\ \emph {et~al.}(2015)\citenamefont {Kurter}, \citenamefont {Finck}, \citenamefont {Hor},\ and\ \citenamefont {Van~Harlingen}}]{kurter_evidence_2015}%
  \BibitemOpen
  \bibfield  {author} {\bibinfo {author} {\bibfnamefont {C.}~\bibnamefont {Kurter}}, \bibinfo {author} {\bibfnamefont {A.~D.~K.}\ \bibnamefont {Finck}}, \bibinfo {author} {\bibfnamefont {Y.~S.}\ \bibnamefont {Hor}},\ and\ \bibinfo {author} {\bibfnamefont {D.~J.}\ \bibnamefont {Van~Harlingen}},\ }\href {https://doi.org/10.1038/ncomms8130} {\bibfield  {journal} {\bibinfo  {journal} {Nature Communications}\ }\textbf {\bibinfo {volume} {6}},\ \bibinfo {pages} {7130} (\bibinfo {year} {2015})}\BibitemShut {NoStop}%
\bibitem [{\citenamefont {Charpentier}\ \emph {et~al.}(2017)\citenamefont {Charpentier}, \citenamefont {Galletti}, \citenamefont {Kunakova}, \citenamefont {Arpaia}, \citenamefont {Song}, \citenamefont {Baghdadi}, \citenamefont {Wang}, \citenamefont {Kalaboukhov}, \citenamefont {Olsson}, \citenamefont {Tafuri}, \citenamefont {Golubev}, \citenamefont {Linder}, \citenamefont {Bauch},\ and\ \citenamefont {Lombardi}}]{charpentier_induced_2017}%
  \BibitemOpen
  \bibfield  {author} {\bibinfo {author} {\bibfnamefont {S.}~\bibnamefont {Charpentier}}, \bibinfo {author} {\bibfnamefont {L.}~\bibnamefont {Galletti}}, \bibinfo {author} {\bibfnamefont {G.}~\bibnamefont {Kunakova}}, \bibinfo {author} {\bibfnamefont {R.}~\bibnamefont {Arpaia}}, \bibinfo {author} {\bibfnamefont {Y.}~\bibnamefont {Song}}, \bibinfo {author} {\bibfnamefont {R.}~\bibnamefont {Baghdadi}}, \bibinfo {author} {\bibfnamefont {S.~M.}\ \bibnamefont {Wang}}, \bibinfo {author} {\bibfnamefont {A.}~\bibnamefont {Kalaboukhov}}, \bibinfo {author} {\bibfnamefont {E.}~\bibnamefont {Olsson}}, \bibinfo {author} {\bibfnamefont {F.}~\bibnamefont {Tafuri}}, \bibinfo {author} {\bibfnamefont {D.}~\bibnamefont {Golubev}}, \bibinfo {author} {\bibfnamefont {J.}~\bibnamefont {Linder}}, \bibinfo {author} {\bibfnamefont {T.}~\bibnamefont {Bauch}},\ and\ \bibinfo {author} {\bibfnamefont {F.}~\bibnamefont {Lombardi}},\ }\href {https://doi.org/10.1038/s41467-017-02069-z} {\bibfield  {journal} {\bibinfo  {journal} {Nature
  Communications}\ }\textbf {\bibinfo {volume} {8}},\ \bibinfo {pages} {2019} (\bibinfo {year} {2017})}\BibitemShut {NoStop}%
\bibitem [{\citenamefont {Ghatak}\ \emph {et~al.}(2018)\citenamefont {Ghatak}, \citenamefont {Breunig}, \citenamefont {Yang}, \citenamefont {Wang}, \citenamefont {Taskin},\ and\ \citenamefont {Ando}}]{ghatak_anomalous_2018}%
  \BibitemOpen
  \bibfield  {author} {\bibinfo {author} {\bibfnamefont {S.}~\bibnamefont {Ghatak}}, \bibinfo {author} {\bibfnamefont {O.}~\bibnamefont {Breunig}}, \bibinfo {author} {\bibfnamefont {F.}~\bibnamefont {Yang}}, \bibinfo {author} {\bibfnamefont {Z.}~\bibnamefont {Wang}}, \bibinfo {author} {\bibfnamefont {A.~A.}\ \bibnamefont {Taskin}},\ and\ \bibinfo {author} {\bibfnamefont {Y.}~\bibnamefont {Ando}},\ }\href {https://doi.org/10.1021/acs.nanolett.8b02029} {\bibfield  {journal} {\bibinfo  {journal} {Nano Letters}\ }\textbf {\bibinfo {volume} {18}},\ \bibinfo {pages} {5124} (\bibinfo {year} {2018})}\BibitemShut {NoStop}%
\bibitem [{\citenamefont {Kayyalha}\ \emph {et~al.}(2019)\citenamefont {Kayyalha}, \citenamefont {Kargarian}, \citenamefont {Kazakov}, \citenamefont {Miotkowski}, \citenamefont {Galitski}, \citenamefont {Yakovenko}, \citenamefont {Rokhinson},\ and\ \citenamefont {Chen}}]{kayyalha_anomalous_2019}%
  \BibitemOpen
  \bibfield  {author} {\bibinfo {author} {\bibfnamefont {M.}~\bibnamefont {Kayyalha}}, \bibinfo {author} {\bibfnamefont {M.}~\bibnamefont {Kargarian}}, \bibinfo {author} {\bibfnamefont {A.}~\bibnamefont {Kazakov}}, \bibinfo {author} {\bibfnamefont {I.}~\bibnamefont {Miotkowski}}, \bibinfo {author} {\bibfnamefont {V.~M.}\ \bibnamefont {Galitski}}, \bibinfo {author} {\bibfnamefont {V.~M.}\ \bibnamefont {Yakovenko}}, \bibinfo {author} {\bibfnamefont {L.~P.}\ \bibnamefont {Rokhinson}},\ and\ \bibinfo {author} {\bibfnamefont {Y.~P.}\ \bibnamefont {Chen}},\ }\href {https://doi.org/10.1103/PhysRevLett.122.047003} {\bibfield  {journal} {\bibinfo  {journal} {Physical Review Letters}\ }\textbf {\bibinfo {volume} {122}},\ \bibinfo {pages} {047003} (\bibinfo {year} {2019})}\BibitemShut {NoStop}%
\bibitem [{\citenamefont {Kayyalha}\ \emph {et~al.}(2020)\citenamefont {Kayyalha}, \citenamefont {Kazakov}, \citenamefont {Miotkowski}, \citenamefont {Khlebnikov}, \citenamefont {Rokhinson},\ and\ \citenamefont {Chen}}]{kayyalha_highly_2020}%
  \BibitemOpen
  \bibfield  {author} {\bibinfo {author} {\bibfnamefont {M.}~\bibnamefont {Kayyalha}}, \bibinfo {author} {\bibfnamefont {A.}~\bibnamefont {Kazakov}}, \bibinfo {author} {\bibfnamefont {I.}~\bibnamefont {Miotkowski}}, \bibinfo {author} {\bibfnamefont {S.}~\bibnamefont {Khlebnikov}}, \bibinfo {author} {\bibfnamefont {L.~P.}\ \bibnamefont {Rokhinson}},\ and\ \bibinfo {author} {\bibfnamefont {Y.~P.}\ \bibnamefont {Chen}},\ }\href {https://doi.org/10.1038/s41535-020-0209-5} {\bibfield  {journal} {\bibinfo  {journal} {npj Quantum Materials}\ }\textbf {\bibinfo {volume} {5}},\ \bibinfo {pages} {1} (\bibinfo {year} {2020})}\BibitemShut {NoStop}%
\bibitem [{\citenamefont {Takeshige}\ \emph {et~al.}(2020)\citenamefont {Takeshige}, \citenamefont {Matsuo}, \citenamefont {Deacon}, \citenamefont {Ueda}, \citenamefont {Sato}, \citenamefont {Zhao}, \citenamefont {Zhou}, \citenamefont {Chang}, \citenamefont {Ishibashi},\ and\ \citenamefont {Tarucha}}]{takeshige_experimental_2020}%
  \BibitemOpen
  \bibfield  {author} {\bibinfo {author} {\bibfnamefont {Y.}~\bibnamefont {Takeshige}}, \bibinfo {author} {\bibfnamefont {S.}~\bibnamefont {Matsuo}}, \bibinfo {author} {\bibfnamefont {R.~S.}\ \bibnamefont {Deacon}}, \bibinfo {author} {\bibfnamefont {K.}~\bibnamefont {Ueda}}, \bibinfo {author} {\bibfnamefont {Y.}~\bibnamefont {Sato}}, \bibinfo {author} {\bibfnamefont {Y.-F.}\ \bibnamefont {Zhao}}, \bibinfo {author} {\bibfnamefont {L.}~\bibnamefont {Zhou}}, \bibinfo {author} {\bibfnamefont {C.-Z.}\ \bibnamefont {Chang}}, \bibinfo {author} {\bibfnamefont {K.}~\bibnamefont {Ishibashi}},\ and\ \bibinfo {author} {\bibfnamefont {S.}~\bibnamefont {Tarucha}},\ }\href {https://doi.org/10.1103/PhysRevB.101.115410} {\bibfield  {journal} {\bibinfo  {journal} {Physical Review B}\ }\textbf {\bibinfo {volume} {101}},\ \bibinfo {pages} {115410} (\bibinfo {year} {2020})}\BibitemShut {NoStop}%
\bibitem [{\citenamefont {Hadfield}\ \emph {et~al.}(2003)\citenamefont {Hadfield}, \citenamefont {Burnell}, \citenamefont {Kang}, \citenamefont {Bell},\ and\ \citenamefont {Blamire}}]{hadfield_corbino_2003}%
  \BibitemOpen
  \bibfield  {author} {\bibinfo {author} {\bibfnamefont {R.~H.}\ \bibnamefont {Hadfield}}, \bibinfo {author} {\bibfnamefont {G.}~\bibnamefont {Burnell}}, \bibinfo {author} {\bibfnamefont {D.-J.}\ \bibnamefont {Kang}}, \bibinfo {author} {\bibfnamefont {C.}~\bibnamefont {Bell}},\ and\ \bibinfo {author} {\bibfnamefont {M.~G.}\ \bibnamefont {Blamire}},\ }\href {https://doi.org/10.1103/PhysRevB.67.144513} {\bibfield  {journal} {\bibinfo  {journal} {Physical Review B}\ }\textbf {\bibinfo {volume} {67}},\ \bibinfo {pages} {144513} (\bibinfo {year} {2003})}\BibitemShut {NoStop}%
\bibitem [{\citenamefont {Clem}(2010)}]{clem_corbino-geometry_2010}%
  \BibitemOpen
  \bibfield  {author} {\bibinfo {author} {\bibfnamefont {J.~R.}\ \bibnamefont {Clem}},\ }\href {https://doi.org/10.1103/PhysRevB.82.174515} {\bibfield  {journal} {\bibinfo  {journal} {Physical Review B}\ }\textbf {\bibinfo {volume} {82}},\ \bibinfo {pages} {174515} (\bibinfo {year} {2010})}\BibitemShut {NoStop}%
\bibitem [{\citenamefont {Matsuo}\ \emph {et~al.}(2020)\citenamefont {Matsuo}, \citenamefont {Tateno}, \citenamefont {Sato}, \citenamefont {Ueda}, \citenamefont {Takeshige}, \citenamefont {Kamata}, \citenamefont {Lee}, \citenamefont {Shojaei}, \citenamefont {Palmstr{\o}m},\ and\ \citenamefont {Tarucha}}]{matsuo_evaluation_2020}%
  \BibitemOpen
  \bibfield  {author} {\bibinfo {author} {\bibfnamefont {S.}~\bibnamefont {Matsuo}}, \bibinfo {author} {\bibfnamefont {M.}~\bibnamefont {Tateno}}, \bibinfo {author} {\bibfnamefont {Y.}~\bibnamefont {Sato}}, \bibinfo {author} {\bibfnamefont {K.}~\bibnamefont {Ueda}}, \bibinfo {author} {\bibfnamefont {Y.}~\bibnamefont {Takeshige}}, \bibinfo {author} {\bibfnamefont {H.}~\bibnamefont {Kamata}}, \bibinfo {author} {\bibfnamefont {J.~S.}\ \bibnamefont {Lee}}, \bibinfo {author} {\bibfnamefont {B.}~\bibnamefont {Shojaei}}, \bibinfo {author} {\bibfnamefont {C.~J.}\ \bibnamefont {Palmstr{\o}m}},\ and\ \bibinfo {author} {\bibfnamefont {S.}~\bibnamefont {Tarucha}},\ }\href {https://doi.org/10.1103/PhysRevB.102.045301} {\bibfield  {journal} {\bibinfo  {journal} {Physical Review B}\ }\textbf {\bibinfo {volume} {102}},\ \bibinfo {pages} {045301} (\bibinfo {year} {2020})}\BibitemShut {NoStop}%
\bibitem [{\citenamefont {Dominguez}\ \emph {et~al.}(2022)\citenamefont {Dominguez}, \citenamefont {Novik},\ and\ \citenamefont {Recher}}]{dominguez_fraunhofer_2022}%
  \BibitemOpen
  \bibfield  {author} {\bibinfo {author} {\bibfnamefont {F.}~\bibnamefont {Dominguez}}, \bibinfo {author} {\bibfnamefont {E.~G.}\ \bibnamefont {Novik}},\ and\ \bibinfo {author} {\bibfnamefont {P.}~\bibnamefont {Recher}},\ }\href {https://doi.org/10.48550/arXiv.2210.02065} {\bibinfo {title} {Fraunhofer pattern in the presence of {{Majorana}} zero modes}} (\bibinfo {year} {2022}),\ \Eprint {https://arxiv.org/abs/2210.02065} {arXiv:2210.02065 [cond-mat]} \BibitemShut {NoStop}%
\bibitem [{\citenamefont {Shapiro}(1963)}]{shapiro_josephson_1963}%
  \BibitemOpen
  \bibfield  {author} {\bibinfo {author} {\bibfnamefont {S.}~\bibnamefont {Shapiro}},\ }\href {https://doi.org/10.1103/PhysRevLett.11.80} {\bibfield  {journal} {\bibinfo  {journal} {Physical Review Letters}\ }\textbf {\bibinfo {volume} {11}},\ \bibinfo {pages} {80} (\bibinfo {year} {1963})}\BibitemShut {NoStop}%
\bibitem [{\citenamefont {Nielsen}\ and\ \citenamefont {Ninomiya}(1981{\natexlab{a}})}]{nielsen_absence_1981}%
  \BibitemOpen
  \bibfield  {author} {\bibinfo {author} {\bibfnamefont {H.~B.}\ \bibnamefont {Nielsen}}\ and\ \bibinfo {author} {\bibfnamefont {M.}~\bibnamefont {Ninomiya}},\ }\href {https://doi.org/10.1016/0550-3213(81)90524-1} {\bibfield  {journal} {\bibinfo  {journal} {Nuclear Physics B}\ }\textbf {\bibinfo {volume} {193}},\ \bibinfo {pages} {173} (\bibinfo {year} {1981}{\natexlab{a}})}\BibitemShut {NoStop}%
\bibitem [{\citenamefont {Nielsen}\ and\ \citenamefont {Ninomiya}(1981{\natexlab{b}})}]{nielsen_absence_1981-1}%
  \BibitemOpen
  \bibfield  {author} {\bibinfo {author} {\bibfnamefont {H.~B.}\ \bibnamefont {Nielsen}}\ and\ \bibinfo {author} {\bibfnamefont {M.}~\bibnamefont {Ninomiya}},\ }\href {https://doi.org/10.1016/0550-3213(81)90361-8} {\bibfield  {journal} {\bibinfo  {journal} {Nuclear Physics B}\ }\textbf {\bibinfo {volume} {185}},\ \bibinfo {pages} {20} (\bibinfo {year} {1981}{\natexlab{b}})}\BibitemShut {NoStop}%
\bibitem [{\citenamefont {Grover}\ \emph {et~al.}(2014)\citenamefont {Grover}, \citenamefont {Sheng},\ and\ \citenamefont {Vishwanath}}]{grover_emergent_2014}%
  \BibitemOpen
  \bibfield  {author} {\bibinfo {author} {\bibfnamefont {T.}~\bibnamefont {Grover}}, \bibinfo {author} {\bibfnamefont {D.~N.}\ \bibnamefont {Sheng}},\ and\ \bibinfo {author} {\bibfnamefont {A.}~\bibnamefont {Vishwanath}},\ }\href {https://doi.org/10.1126/science.1248253} {\bibfield  {journal} {\bibinfo  {journal} {Science}\ }\textbf {\bibinfo {volume} {344}},\ \bibinfo {pages} {280} (\bibinfo {year} {2014})}\BibitemShut {NoStop}%
\bibitem [{\citenamefont {Li}\ \emph {et~al.}(2020)\citenamefont {Li}, \citenamefont {Ebisu}, \citenamefont {Sahoo}, \citenamefont {Oreg},\ and\ \citenamefont {Franz}}]{li_coupled_2020}%
  \BibitemOpen
  \bibfield  {author} {\bibinfo {author} {\bibfnamefont {C.}~\bibnamefont {Li}}, \bibinfo {author} {\bibfnamefont {H.}~\bibnamefont {Ebisu}}, \bibinfo {author} {\bibfnamefont {S.}~\bibnamefont {Sahoo}}, \bibinfo {author} {\bibfnamefont {Y.}~\bibnamefont {Oreg}},\ and\ \bibinfo {author} {\bibfnamefont {M.}~\bibnamefont {Franz}},\ }\href {https://doi.org/10.1103/PhysRevB.102.165123} {\bibfield  {journal} {\bibinfo  {journal} {Physical Review B}\ }\textbf {\bibinfo {volume} {102}},\ \bibinfo {pages} {165123} (\bibinfo {year} {2020})}\BibitemShut {NoStop}%
\bibitem [{\citenamefont {Setiawan}\ \emph {et~al.}(2019)\citenamefont {Setiawan}, \citenamefont {Stern},\ and\ \citenamefont {Berg}}]{setiawan_topological_2019}%
  \BibitemOpen
  \bibfield  {author} {\bibinfo {author} {\bibfnamefont {F.}~\bibnamefont {Setiawan}}, \bibinfo {author} {\bibfnamefont {A.}~\bibnamefont {Stern}},\ and\ \bibinfo {author} {\bibfnamefont {E.}~\bibnamefont {Berg}},\ }\href {https://doi.org/10.1103/PhysRevB.99.220506} {\bibfield  {journal} {\bibinfo  {journal} {Physical Review B}\ }\textbf {\bibinfo {volume} {99}},\ \bibinfo {pages} {220506} (\bibinfo {year} {2019})}\BibitemShut {NoStop}%
\bibitem [{Sup()}]{SupplementalMaterial}%
  \BibitemOpen
  \href@noop {} {}\bibinfo {note} {See Supplemental Material for details on the connection between period halving and the current-phase relation, and further elaboration on the square-like geometry and low-lying wavefunctions of the tight-binding model.}\BibitemShut {Stop}%
\bibitem [{\citenamefont {Park}\ \emph {et~al.}(2026)\citenamefont {Park}, \citenamefont {Werkmeister}, \citenamefont {Zauberman}, \citenamefont {Lesser}, \citenamefont {Anderson}, \citenamefont {Ronen}, \citenamefont {Medina~Cea}, \citenamefont {Kushwaha}, \citenamefont {Watanabe}, \citenamefont {Taniguchi}, \citenamefont {Cava}, \citenamefont {Oreg}, \citenamefont {Yacoby},\ and\ \citenamefont {Kim}}]{park_2025}%
  \BibitemOpen
  \bibfield  {author} {\bibinfo {author} {\bibfnamefont {J.~Y.}\ \bibnamefont {Park}}, \bibinfo {author} {\bibfnamefont {T.}~\bibnamefont {Werkmeister}}, \bibinfo {author} {\bibfnamefont {J.}~\bibnamefont {Zauberman}}, \bibinfo {author} {\bibfnamefont {O.}~\bibnamefont {Lesser}}, \bibinfo {author} {\bibfnamefont {L.~E.}\ \bibnamefont {Anderson}}, \bibinfo {author} {\bibfnamefont {Y.}~\bibnamefont {Ronen}}, \bibinfo {author} {\bibfnamefont {C.~J.}\ \bibnamefont {Medina~Cea}}, \bibinfo {author} {\bibfnamefont {S.~K.}\ \bibnamefont {Kushwaha}}, \bibinfo {author} {\bibfnamefont {K.}~\bibnamefont {Watanabe}}, \bibinfo {author} {\bibfnamefont {T.}~\bibnamefont {Taniguchi}}, \bibinfo {author} {\bibfnamefont {R.~J.}\ \bibnamefont {Cava}}, \bibinfo {author} {\bibfnamefont {Y.}~\bibnamefont {Oreg}}, \bibinfo {author} {\bibfnamefont {A.}~\bibnamefont {Yacoby}},\ and\ \bibinfo {author} {\bibfnamefont {P.}~\bibnamefont {Kim}},\ }\href@noop {} {\bibfield  {journal} {\bibinfo  {journal} {preprint}\ } (\bibinfo {year}
  {2026})}\BibitemShut {NoStop}%
\bibitem [{\citenamefont {Kushwaha}\ \emph {et~al.}(2016)\citenamefont {Kushwaha}, \citenamefont {Pletikosi{\'c}}, \citenamefont {Liang}, \citenamefont {Gyenis}, \citenamefont {Lapidus}, \citenamefont {Tian}, \citenamefont {Zhao}, \citenamefont {Burch}, \citenamefont {Lin}, \citenamefont {Wang}, \citenamefont {Ji}, \citenamefont {Fedorov}, \citenamefont {Yazdani}, \citenamefont {Ong}, \citenamefont {Valla},\ and\ \citenamefont {Cava}}]{kushwaha_sn-doped_2016}%
  \BibitemOpen
  \bibfield  {author} {\bibinfo {author} {\bibfnamefont {S.~K.}\ \bibnamefont {Kushwaha}}, \bibinfo {author} {\bibfnamefont {I.}~\bibnamefont {Pletikosi{\'c}}}, \bibinfo {author} {\bibfnamefont {T.}~\bibnamefont {Liang}}, \bibinfo {author} {\bibfnamefont {A.}~\bibnamefont {Gyenis}}, \bibinfo {author} {\bibfnamefont {S.~H.}\ \bibnamefont {Lapidus}}, \bibinfo {author} {\bibfnamefont {Y.}~\bibnamefont {Tian}}, \bibinfo {author} {\bibfnamefont {H.}~\bibnamefont {Zhao}}, \bibinfo {author} {\bibfnamefont {K.~S.}\ \bibnamefont {Burch}}, \bibinfo {author} {\bibfnamefont {J.}~\bibnamefont {Lin}}, \bibinfo {author} {\bibfnamefont {W.}~\bibnamefont {Wang}}, \bibinfo {author} {\bibfnamefont {H.}~\bibnamefont {Ji}}, \bibinfo {author} {\bibfnamefont {A.~V.}\ \bibnamefont {Fedorov}}, \bibinfo {author} {\bibfnamefont {A.}~\bibnamefont {Yazdani}}, \bibinfo {author} {\bibfnamefont {N.~P.}\ \bibnamefont {Ong}}, \bibinfo {author} {\bibfnamefont {T.}~\bibnamefont {Valla}},\ and\ \bibinfo {author} {\bibfnamefont {R.~J.}\
  \bibnamefont {Cava}},\ }\href {https://doi.org/10.1038/ncomms11456} {\bibfield  {journal} {\bibinfo  {journal} {Nature Communications}\ }\textbf {\bibinfo {volume} {7}},\ \bibinfo {pages} {11456} (\bibinfo {year} {2016})}\BibitemShut {NoStop}%
\end{thebibliography}%

\end{document}


\newcommand{\myTitle}{Theory of reentrant superconductivity in Corbino Josephson junctions}

\newcommand{\Weizmann}{Department of Condensed Matter Physics, Weizmann Institute of Science, Rehovot, Israel 7610001}
\newcommand{\Cornell}{Department of Physics, Cornell University, Ithaca, NY 14853, USA}
\newcommand{\HarvardPhysics}{Department of Physics, Harvard University, Cambridge, Massachusetts 02138, USA}
\newcommand{\HarvardEngineering}{John A. Paulson School of Engineering and Applied Sciences, Harvard University, Cambridge, Massachusetts 02138, USA}
\newcommand{\Columbia}{Department of Applied Physics and Applied Mathematics, Columbia University, New York, New York 10027, USA}
\newcommand{\SKKU}{Department of Physics, Sungkyunkwan University (SKKU), Suwon 16419, Republic of Korea}
\newcommand{\IBS}{Center for 2D Quantum Heterostructures, Institute for Basic Science (IBS), Sungkyunkwan University (SKKU), Suwon 16419, Republic of Korea}

\author{Omri Lesser}
\affiliation{\Cornell}
\affiliation{\Weizmann}

\author{Joon Young Park}
\affiliation{\HarvardPhysics}
\affiliation{\SKKU}
\affiliation{\IBS}

\author{Yuval Ronen}
\affiliation{\Weizmann}

\author{Thomas Werkmeister}
\affiliation{\HarvardEngineering}
\affiliation{\Columbia}

\author{Philip Kim}
\affiliation{\HarvardPhysics}
\affiliation{\HarvardEngineering}

\author{Yuval Oreg}
\affiliation{\Weizmann}
\title{Supplemental Material for\\*[0.5em]``\myTitle"}
\maketitle

\setcounter{equation}{0}
\renewcommand{\theequation}{S\arabic{equation}}
\setcounter{figure}{0}
\renewcommand{\thefigure}{S\arabic{figure}}
\setcounter{section}{0}
\renewcommand{\thesection}{S\Roman{section}}
\onecolumngrid

\section{Period halving from current-phase relation}
In this section, we show that the period halving observed numerically for non-circular topological junctions is consistent with a high-frequency term in the current-phase relation.
To do this, we consider the general current-phase relation
\begin{equation}
    I\left(\phi\right)=\sum_{k=1}^{\infty}I_{0}^{\left(k\right)}\sin\left(k\phi\right).
\end{equation}
In conventional junctions, we generally expect $I_{0}^{\left(k\right)}=0$ for $k\neq 1$. Plugging in the simple harmonic form of the phase evolution $\phi(\theta) = n_{\rm v}\theta + a\sin(n_{\rm c}\theta)$ [Eq.~\eqref{eq:single_harmony} of the main text], we calculate the critical current as
\begin{equation}
    \begin{aligned}I_{c} & =\max_{\phi_{0}}\sum_{k=1}^{\infty}I_{0}^{\left(k\right)}\int_{0}^{2\pi}d\theta\sin\left[k\phi_{0}+kn_{{\rm v}}\theta+ka\sin\left(n_{\text{c}}\theta\right)\right]\\
 & =\max_{\phi_{0}}\sum_{k=1}^{\infty}I_{0}^{\left(k\right)}\sum_{m=0}^{\infty}\left(-1\right)^{m}J_{m}\left(ka\right)\int_{0}^{2\pi}d\theta\sin\left[k\phi_{0}+\left(kn_{{\rm v}}-mn_{\text{c}}\right)\theta\right].
\end{aligned}
\end{equation}
This expression is only nonzero if there are integers $k,m$ such that $kn_{{\rm v}}=mn_{\text{c}}$ and $I_{0}^{\left(k\right)}\neq0$.
For the conventional case only $I_{0}^{\left(1\right)}$ is nonzero, and therefore the requirement is that $n_{\rm v}$ is an integer multiple of $n_{\rm c}$. However, if $I_{0}^{\left(2\right)}$ is also nonzero, then the requirement becomes $n_{\rm v}=\frac{m}{2}n_{\rm c}$, i.e., half-integer multiples of $n_{\rm c}$ also yield nonzero critical current. This is exactly what we found numerically for the topological junction, suggesting that this junction has a nonzero $\sin\left(2\phi\right)$ in its current-phase relation. Notice that this is also consistent with the finding that the \emph{circular} topological junction behaves the same as the circular conventional junction.

\section{Geometry and wavefunctions}
In this section we provide additional technical details pertaining to the numerical calculations of the critical current and the tight-binding model.

\begin{figure}[h]
    \centering
   \includegraphics[width=0.6\linewidth]{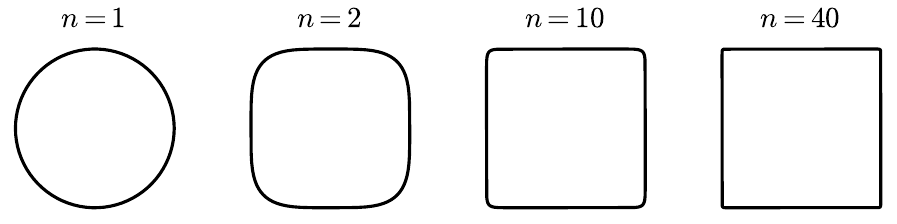}
    \caption{Plots of the implicit equation $x^{2n}+y^{2n}=1$, which is equivalent to the polar equation $r\left(\theta\right)=\left( \cos^{2n}\theta + \sin^{2n}\theta \right)^{-1/2n}$, for $n=1,2,5,20$. This polar form is used to model a square Corbino Josephson junction, with the integer $n$ controlling the sharpness of the corners: $n=1$ is a circle, whereas $n\to\infty$ is a perfect square.
    \label{fig:squares_n}}
\end{figure}

In Fig.~\ref{fig:squares_n} we show the curves used to approximate a square junction. The implicit equation plotted is $x^{2n}+y^{2n}=1$, with $n$ being an integer. At $n=2$ we get a circle, and as $n$ becomes larger the shape becomes more and more square-like. The limit $n\to\infty$ corresponds to perfectly sharp corners. In the numerical calculations we used $n=10$.

\begin{figure}[h]
    \centering
   \includegraphics[width=0.7\linewidth]{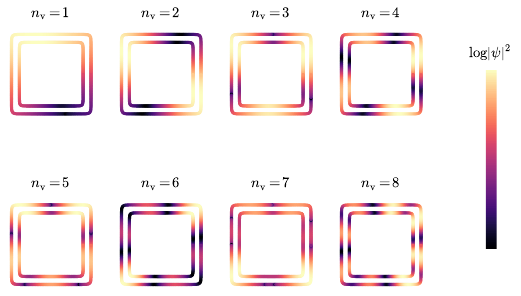}
    \caption{Lowest-energy wavefunctions of the rectangular 3DTI Corbino Josephson junctions, calculated by numerically diagonalizing the tight-binding Hamiltonian [see Eqs.~\eqref{eq:H_ladder}, \eqref{eq:H_Delta} of the main text], for different numbers of vortices $n_{\rm v}$. The wavefunctions are localized around the vortices, i.e., near places where the Josephson coupling vanishes. The wavefunctions are visualized as color plots on top of the inner and outer edges of the normal region separating the two superconductors. The simulations are performed with 400 sites, and the tight-binding parameters are $t_0=1$, $t_1=0.6$, $t_2=0.3$, $\Delta=0.2$.
    \label{fig:wavefunctions_log}}
\end{figure}

In Fig.~\ref{fig:wavefunctions_log} we show the lowest-lying wavefunctions in the tight-binding model of a Corbino JJ on top of a 3DTI. Each additional vortex increases the number of kinks in the wavefunction. The global phase difference $\phi_0$ between the inner and outer superconductors is chosen so as to maximize the Josephson current, and it controls the locations of the kinks.

\begin{figure}[h]
    \centering
   \includegraphics[width=\linewidth]{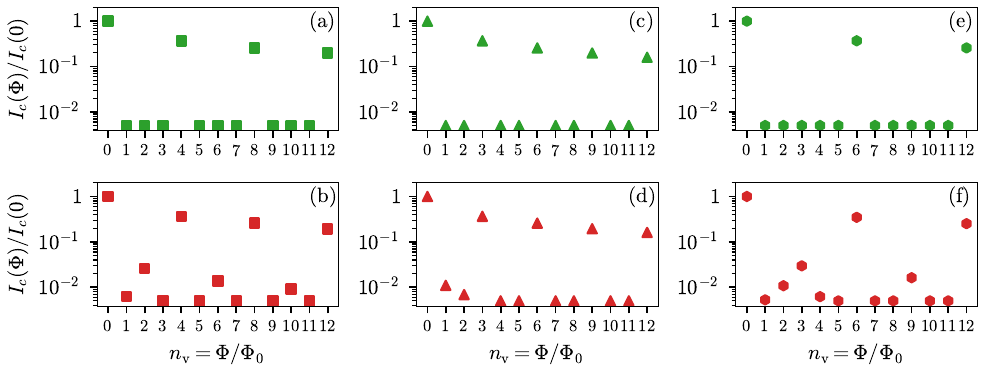}
    \caption{Critical current as a function of the number of vortices $n_{\rm v}$ for (a)~a conventional square Corbino Josephson junction, (b)~a topological square junction, (c)~a conventional triangular junction, (d)~a topological triangular junction, (e)~a conventional hexagonal junction, and (f)~a topological hexagonal junction. Compared to the conventional cases, we observe a period halving in the topological square and hexagonal junctions but not in the topological triangular junction, since in the latter case the number of corners (three) is odd. The simulation parameters are the same as in Fig.~\ref{fig:I_c_all} of the main text. For clarity of the logarithmic scale, we cut off the signals from below at $5\cdot10^{-3}$.
    \label{fig:I_c_shapes}}
\end{figure}

\begin{figure}[h]
    \centering
   \includegraphics[width=\linewidth]{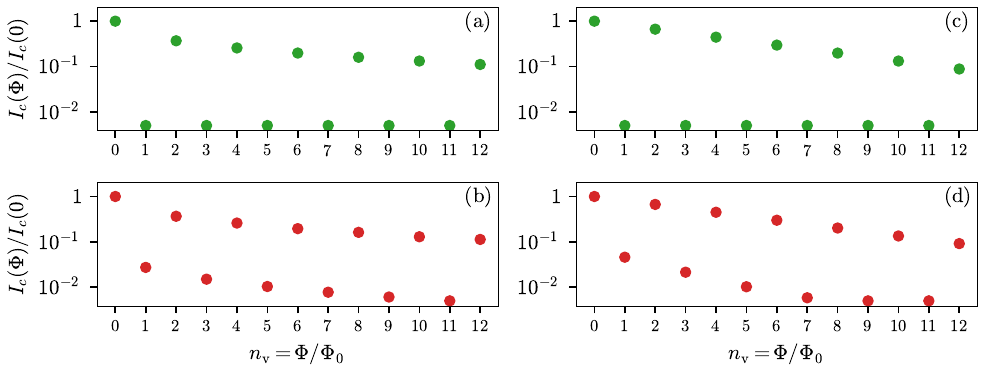}
    \caption{Critical current as a function of the number of vortices $n_{\rm v}$ for Corbino Josephson junctions having effectively two corners. (a) and (b) are the conventional and topological cases, respectively, for a junction with the single-harmony phase approximation of Eq.~\eqref{eq:single_harmony} of the main text with $n_{\rm c}=2$. (c) and (d) are the conventional and topological cases, respectively, for an elliptic junction with eccentricity $e=0.98$. The elliptic junctions mimics the behavior of the $n_{\rm c}=2$ junction: in the conventional cases superconductivity only appears at even $n_{\rm v}$, whereas in the topological case superconductivity is diminished but non-vanishing also at odd $n_{\rm v}$. The simulation parameters are the same as in Fig.~\ref{fig:I_c_all} of the main text. For clarity of the logarithmic scale, we cut off the signals from below at $5\cdot10^{-3}$.
    \label{fig:I_c_two_corners}}
\end{figure}

In Fig.~\ref{fig:I_c_shapes} we show the critical current as a function of the number of vortices for square, triangular, and hexagonal Corbino Josephson junctions. These three shapes are modeled according to Eq.~\eqref{eq:single_harmony} of the main text, with $n_{\rm c}=4,3,6$ respectively. In the conventional junctions, we observe reentrant superconductivity at when $n_{\rm v}$ is an integer multiple of $n_{\rm c}$, as expected. In the topological case, we find different types of behavior: in the square and hexagonal cases superconductivity reenters also when $n_{\rm v}$ is a half-integer multiple of $n_{\rm c}$. However, in the triangular case, $n_{\rm c}/2=3/2$ is not an integer, and therefore superconductivity cannot reenter at half-integer multiples of $n_{\rm c}$. Therefore, in the triangular case we find no qualitative difference between the conventional and topological cases.
Fig.~\ref{fig:I_c_two_corners} shows the analogous plots for junctions have effectively two corners, one with the explicit single-harmony approximation of Eq.~\eqref{eq:single_harmony} of the main text with $n_{\rm c}=2$, and one with an elliptic geometry. In this case, we find that the conventional junctions has a nonzero critical current only for even $n_{\rm v}$. On the other hand, the topological junctions have nonzero (yet reduced) critical current also at odd $n_{\rm v}$.

